\documentclass[aps,twocolumn,preprintnumbers,showpacs,showkeys,nofootinbib]{revtex4-1}
\usepackage{epsfig,float,amssymb,latexsym,amsmath,amsthm,fancyhdr}
\usepackage{graphics,longtable}
\newcommand{\ba}{\begin{eqnarray}}
\newcommand{\ea}{\end{eqnarray}}
\newcommand{\be}{\begin{equation}}
\newcommand{\ee}{\end{equation}}
\begin{document}
\author{J.~M.~Alarc\'on}
\email{alarcon@jlab.org}
\affiliation{Theory Center, Jefferson Lab, Newport News, VA 23606, USA}
\author{C.~Weiss} 
\email{weiss@jlab.org}
\affiliation{Theory Center, Jefferson Lab, Newport News, VA 23606, USA}
\preprint{JLAB-THY-17-2568}
\title{Nucleon form factors in dispersively improved Chiral Effective Field Theory II: \\
Electromagnetic form factors}
\begin{abstract}
We study the nucleon electromagnetic form factors (EM FFs) using a recently developed method
combining Chiral Effective Field Theory ($\chi$EFT) and dispersion analysis. The spectral functions
on the two-pion cut at $t > 4 M_\pi^2$ are constructed using the elastic unitarity relation 
and an $N/D$ representation. $\chi$EFT is used to calculate the real functions 
$J_\pm^1 (t) = f_\pm^1(t)/F_\pi(t)$ (ratios of the complex $\pi\pi \rightarrow N \bar N$ 
partial-wave amplitudes and the timelike pion FF), which are free of $\pi\pi$ rescattering. 
Rescattering effects are included through the empirical timelike pion FF $|F_\pi(t)|^2$. 
The method allows us to compute the isovector EM spectral functions up to $t \sim 1$ GeV$^2$ 
with controlled accuracy (LO, NLO, and partial N2LO). With the spectral functions we calculate 
the isovector nucleon EM FFs and their derivatives at $t = 0$ (EM radii, moments) using subtracted 
dispersion relations. We predict the values of higher FF derivatives with minimal uncertainties 
and explain their collective behavior. We estimate the individual proton and neutron FFs by adding 
an empirical parametrization of the isoscalar sector. Excellent agreement with the present 
low-$Q^2$ FF data is achieved up to $\sim$0.5 GeV$^2$ for $G_E$, and up to $\sim$0.2 GeV$^2$ 
for $G_M$. Our results can be used to guide the analysis of low-$Q^2$ 
elastic scattering data and the extraction of the proton charge radius.
\end{abstract}
\maketitle
\tableofcontents
\section{Introduction}
The electromagnetic form factors (EM FFs) describe the nucleon's elastic response to external 
EM fields and reveal the spatial distribution of charge and magnetization inside the strongly
interacting system. They are among the most basic characteristics of nucleon structure and have
been the object of extensive theoretical and experimental study. The electric and magnetic FFs
at invariant momentum transfers $t < 0$ are measured in elastic electron-nucleon scattering.
Experiments at $|t| \leq 1\, \textrm{GeV}^2$ have been performed at many facilities, 
most recently at the Mainz Microtron (MAMI) \cite{Bernauer:2010wm,Bernauer:2013tpr,Mihovilovic:2016rkr}
and at Jefferson Lab (JLab) \cite{Crawford:2006rz,Paolone:2010qc,Zhan:2011ji}; 
see Refs.~\cite{Perdrisat:2006hj,Punjabi:2015bba} for a review of the other data.
The derivative of the proton electric FF at $t = 0$, or charge radius, governs the nucleon structure
corrections to the energy levels of hydrogen atoms (electronic or muonic) and is measured with 
high precision in atomic physics experiments. Recent experimental results have raised interesting 
questions regarding the precise value of the proton charge radius and the extrapolation of the 
elastic scattering data to $t \rightarrow 0$; see Refs.~\cite{Pohl:2010zza,Carlson:2015jba,Krauth:2017ijq} 
for a review. Theoretical calculations of the nucleon FFs at $|t| \ll 1$ GeV$^2$ are needed 
to guide the analysis of the experimental data and help answer these questions. Dedicated measurements 
of the proton electric FF at extremely low momentum transfers $|t| \geq 10^{-4} \, \textrm{GeV}^2$ 
are planned at JLab \cite{Gasparian:2017cgp}. Knowledge of the low-$t$ EM FFs is also required
for constructing the peripheral transverse densities and generalized 
parton distributions (GPDs) of the nucleon \cite{Granados:2013moa,Alarcon:2017asr}.

In a previous article we have proposed a method for calculating the nucleon FFs of $G$-parity-even 
operators by combining Chiral Effective Field Theory ($\chi$EFT) and dispersion analysis
(dispersively improved $\chi$EFT, or DI$\chi$EFT) \cite{Alarcon:2017ivh}. 
It starts from the dispersive representation of the FFs as analytic
functions of $t$ and constructs the spectral functions on the two-pion cut at $t > 4 M_\pi^2$ 
using the elastic unitarity relation \cite{Frazer:1960zza,Frazer:1960zzb}. An $N/D$ representation 
is employed to separate the coupling of the $\pi\pi$ system to nucleon from the effects of $\pi\pi$
rescattering in $t$-channel. $\chi$EFT is used to calculate the real functions $J(t) = f(t)/F_\pi(t)$ ---
the ratios of the complex $\pi\pi \rightarrow N \bar N$ partial-wave amplitudes (PWAs) and the 
timelike pion FF, which are free of $\pi\pi$ rescattering and show good convergence. Rescattering effects 
are included through the timelike pion FF $|F_\pi(t)|^2$, which is taken from sources outside
of $\chi$EFT [experimental data, Lattice QCD (LQCD)]. The formulation permits 
first-principles calculations of the two-pion spectral functions of the FFs with controlled accuracy.
The spectral functions can then be used to evaluate the FFs and related quantities of interest 
(nucleon radii, transverse densities) with subtracted dispersion relations. The method results
in a dramatic improvement compared to conventional $\chi$EFT calculations of nucleon FFs and their 
spectral functions \cite{Gasser:1987rb,Bernard:1996cc,Becher:1999he,Kubis:2000zd,Kaiser:2003qp}.
In Ref.~\cite{Alarcon:2017ivh} the method was applied to the nucleon isoscalar-scalar FF, 
where the $\pi\pi$ system in the $t$--channel is in the $I = J = 0$ state. The resulting
spectral functions and FFs were found to be in good agreement with those of empirical 
dispersion theory and Roy-Steiner equations \cite{Hoehler,Hoferichter:2012wf}.

In the present article we use the DI$\chi$EFT method to calculate the nucleon EM FFs at low $t$ 
and study their properties. We construct the isovector EM spectral functions on the two-pion
cut by combining the elastic unitarity relation in the $I = J = 1$ channel, the $N/D$ representation;
$\chi$EFT calculations of the $J$ functions at LO, NLO and partial N2LO accuracy; and the 
timelike pion FF $|F_\pi|^2$ measured in $e^+e^-$ annihilation experiments. Realistic spectral functions 
are obtained up to $t \sim 1\,\textrm{GeV}^2$, which includes the $\rho$ meson region essential 
for EM structure. With these spectral functions we evaluate the isovector FFs and their derivatives 
(radii) using subtracted dispersion relations. We obtain the individual proton 
and neutron FFs by supplementing the calculated isovector spectral functions with an empirical 
parametrization of the isoscalar ones \cite{Alarcon:2017asr}. Excellent agreement with 
the low-$t$ EM FF data is achieved.

In particular, the method allows us to predict the higher derivatives of the EM FFs at $t = 0$
and explain their collective behavior. They are given by well-convergent dispersion integrals,
which can be evaluated reliably with the DI$\chi$EFT spectral functions, with minimal model
dependence. The higher derivatives are governed by two disparate dynamical scales --- the vector
meson mass, $M_V^2 \; (V = \rho, \omega)$, and the two-pion threshold, $4 M_\pi^2$ --- and exhibit
a rich structure. The values of the higher derivatives therefore differ qualitatively from what 
one would estimate based on a single dynamical scale (naturalness).
Recent attempts to fit the low-$t$ proton electric FF data and extract the charge radius have 
engendered a debate regarding the values of higher FF derivatives and their role in the 
$t \rightarrow 0$ extrapolation \cite{Bernauer:2010wm,Bernauer:2013tpr,Lorenz:2012tm,Lee:2015jqa,%
Higinbotham:2015rja,Griffioen:2015hta,Horbatsch:2016ilr,Sick:2017aor}.
Our predictions for the higher derivatives can be compared with those obtained in
form factor fits (regarding order-of-magnitude, collective behavior) and used
to discriminate between different fits. Our method incorporates the exact 
analytic structure of the FF in $t$ and the multiple dynamical scales governing its
behavior, which are essential in the analysis of low-$t$ FF data and the
extraction of charge radius.

The plan of the article is as follows. In Sec.~\ref{sec:calculation} we describe the steps
of the DI$\chi$EFT calculation, expanding on the general description of the method in 
Ref.~\cite{Alarcon:2017ivh} and emphasizing the aspects that are new or specific to the EM FFs.
This includes the unitarity relations and $N/D$ representation in the $I = J = 1$ channel,
the LO $\chi$EFT calculation of the $J$-functions, the estimate of higher-order
corrections, and the parametrization of the timelike pion FF.
In Sec.~\ref{sec:results} we present the results and their interpretation. This covers
the isovector nucleon EM spectral functions, the nucleon EM radii, higher derivatives (moments) 
of the EM FFs and their structure, and the spacelike nucleon FFs at low $|t|$.
In Sec.~\ref{sec:summary} we summarize the results 
and comment on possible further applications of the method.

A similar method for calculating nucleon FFs, combining $\chi$EFT and dispersion theory, was 
described in Ref.~\cite{Granados:2017cib} and applied to the EM FFs in Ref.~\cite{Leupold:2017ngs}.
The differences from our approach are mainly in the technical implementation of the $\pi\pi$ rescattering 
effects in the unitarity relations ($N/D$ method vs.\ Omnes function) and the estimates of higher-order
chiral corrections. Nucleon FFs were also studied in an extension of $\chi$EFT with explicit vector 
meson degrees of freedom in Ref.~\cite{Bauer:2012pv}. The low--$t$ nucleon FFs and their derivatives
were also calculated in heavy-baryon $\chi$EFT with $\Delta$'s, in the context of a study of 
two-photon exchange corrections to muonic hydrogen in Ref.~\cite{Peset:2014jxa}. 
\section{Calculation}
\label{sec:calculation}
\subsection{Nucleon EM form factors in DI$\chi$EFT}
The transition matrix element of the EM current between nucleon (proton, neutron) states is parametrized 
by the invariant FFs $F_1(t)$ and $F_2(t)$ (Dirac and Pauli FFs; we use the conventions of 
Ref.~\cite{Alarcon:2017asr}). The electric and magnetic FFs are defined as
\ba
G_E (t) &=& F_1(t) - \tau F_2(t), \\
G_M (t) &=& F_1(t) + F_2(t),
\ea
where $\tau \equiv -t/(4 m_N^2)$. At zero momentum transfer the electric FF gives
the electric charge of the nucleon, $G_E^{p, n}(0) = (1, 0)$, and the magnetic FF gives 
the total (pointlike plus anomalous) magnetic moment in units of nuclear magnetons, 
$G_M^{p, n}(0) = \mu^{p, n} = (2.79, -1.91)$. The isovector and isoscalar components
are defined as
\be
G_i^{V, S} \; \equiv \; {\textstyle\frac{1}{2}}(G_i^p \mp G_i^n) \hspace{2em} (i = E, M).
\ee
The FFs are analytic functions of $t$ and satisfy dispersion relations.
They express the FFs at complex $t$ as an integral over their singularities at real 
$t > 0$, corresponding to (unphysical) timelike processes in which the current couples
to the nucleon through exchange of a hadronic system in the $t$-channel. In the case
of the isovector EM FFs the lowest-mass hadronic state is the $\pi\pi$ state, and the 
dispersion integrals start at $t' = 4 M_\pi^2$ (two-pion threshold)
\be
G_i^V (t) \;\; = \;\; \frac{1}{\pi}
\int_{4 M_\pi^2}^\infty dt' \; \frac{\textrm{Im}\, G_i^V (t')}{t' - t - i0} 
\hspace{2em} (i=E, M) .
\label{disp_ff}
\ee
The real functions $\textrm{Im}\, G_i^V (t') \; (t' > 4 M_\pi^2)$ are referred to as 
the spectral functions. At $4 M_\pi^2 < t' < 16 M_\pi^2$ the $\pi\pi$ state is the only 
possible state contributing to $\textrm{Im}\, G_i^V (t')$. It is known from $e^+e^-$
annihilation experiments that higher states ($4\pi$ etc.) do not couple strongly
to the EM current, and it is expected that the $\pi\pi$ channel in the nucleon spectral
functions remains dominant up to $t'\sim 1\,$ GeV$^2$. In this region the $\pi\pi$ channel 
includes the $\rho$ resonance at $t' = M_\rho^2 \sim 0.6 \, \textrm{GeV}^2$, which has a decisive 
influence on the EM FFs. In the isoscalar case the dispersion integral starts with the $3\pi$ state 
at $t' = 9 M_\pi^2$, with the dominant strength at $t< 1\,$ GeV$^2$ coming from the $\omega$ 
resonance at $t' = M_\omega^2 \sim 0.6 \, \textrm{GeV}^2$. The contribution of higher-mass 
states to the isovector and isoscalar FFs is constrained by the total charge and magnetic moment
and has been determined empirically through fits of spacelike FF data
\cite{Hohler:1976ax,Belushkin:2006qa}; the exact composition of these multi-hadron states is 
poorly known and will not be needed in the following applications.

In the region $4 M_\pi^2 < t < 16 M_\pi^2$ the isovector spectral functions on the
two-pion cut can be obtained from the elastic unitarity 
relations \cite{Frazer:1960zza,Frazer:1960zzb}
\ba
\text{Im}\, G_E^V(t) &=& \frac{k_{\rm cm}^3}{m_N \sqrt{t}} \; f_+^1(t) \; F_\pi^*(t) ,
\label{unitarity_G_E}
\\
\text{Im}\, G_M^V(t) &=& \frac{k_{\rm cm}^3}{ \sqrt{2t}} \; f_-^1(t) \; F_\pi^*(t) ,
\label{unitarity_G_M}
\ea
where $k_{\rm cm} \equiv \sqrt{t/4 - M_\pi^2}$ is the center-of-mass momentum of the $\pi\pi$ system
in the $t$-channel, $f_\pm^1(t)$ are the $\pi \pi \rightarrow N\bar{N}$ PWAs
in the normalization of Ref.~\cite{Hohler:1976ax}, and $F_{\pi}^\ast(t)$ is the complex-conjugate 
timelike pion EM FF. Equations~(\ref{unitarity_G_E}) and (\ref{unitarity_G_M}) are valid strictly 
in the region up to the $4\pi$ threshold, $4 \, M_\pi^2 < t < 16 \, M_\pi^2$; if contributions 
from $4\pi$ and higher states are neglected they can effectively be used up to 
$t \sim 1 \textrm{GeV}^2 = 50 \, M_\pi^2$.
The expressions on the right-hand side of Eqs.~(\ref{unitarity_G_E}) and (\ref{unitarity_G_M}) 
are real because the complex functions $f_\pm^1(t)$ and $F_\pi(t)$ have the same phase on the 
two-pion cut (Watson theorem) \cite{Watson:1954uc}. The unitarity relations can therefore be written 
in a manifestly real form as \cite{Frazer:1960zza,Frazer:1960zzb,Hohler:1974eq} 
\ba
\text{Im} \, G_E^V(t) &=& \frac{k_{\rm cm}^3}{m_N \sqrt{t}} \; J_+^1(t) \; |F_\pi(t)|^2 ,
\label{unitarity_G_E_real} 
\\
\text{Im} \, G_M^V(t) &=& \frac{k_{\rm cm}^3}{\sqrt{2t}} \; J_-^1(t) \; |F_\pi(t)|^2 ,
\label{unitarity_G_M_real} 
\ea
where
\be
J_\pm^1(t) \;\; \equiv \;\; \frac{f_\pm^1(t)}{F_\pi(t)}.
\label{J_pm_def}
\ee
The functions $J_\pm^1(t)$ are real for $t > 4 M_\pi^2$ and thus have no right-hand cut;
their only singularities are left-hand cuts at $t < 4 M_\pi^2 - M_\pi^4/m_N^2$, the threshold 
resulting from the singularity of the nucleon Born term in the $\pi\pi \rightarrow N\bar N$
PWAs. Equations~(\ref{unitarity_G_E_real})--(\ref{J_pm_def})
are equivalent to a particular $N/D$ representation of the PWAs \cite{Chew:1960iv},
\be
f_\pm^1(t) \;\; = \;\; \frac{J_\pm^1 (t)}{D(t)},
\hspace{2em} D(t) \equiv \; 1/F_\pi(t) ,
\label{N_over_D}
\ee
in which the numerator functions $J_\pm^1(t)$ contain the left-hand cut and the
denominator function $1/F_\pi(t)$ contains the right-hand cut.

To evaluate the spectral functions in the representation of 
Eqs.~(\ref{unitarity_G_E_real})--(\ref{J_pm_def}), following Ref.~\cite{Alarcon:2017ivh}, 
we calculate the real functions $J_\pm^1(t)$ in $\chi$EFT and multiply 
them with the empirical pion FF modulus $|F_\pi|^2$.
Advantages of this approach are: (a) The $\chi$EFT calculation of $J_\pm^1(t)$ is free of $\pi\pi$ 
rescattering and shows good convergence. Rescattering effects are entirely contained 
in $|F_\pi(t)|^2$, which is taken from other sources. In traditional ``direct'' $\chi$EFT calculations
of the spectral functions the $\pi\pi$ rescattering effects result in large higher-order corrections
and render the perturbative expansion impractical. (b) The functions $J_\pm^1(t)$ are dominated 
by the scales $M_\pi$ and $m_\Delta - m_N$ associated with the Born graph singularities, while
$|F_\pi(t)|^2$ dominated by the chiral-symmetry-breaking scale $\Lambda_\chi$. The representation 
Eqs.~(\ref{unitarity_G_E_real})--(\ref{J_pm_def}) is therefore consistent with the 
idea of separation of scales. (c) The squared modulus $|F_\pi(t)|^2$ can be imported directly 
from the $e^+e^- \rightarrow \pi^+\pi^-$ data or from LQCD calculations without determination 
of the phase \cite{Alarcon:2017ivh,Meyer:2011um}.
For further discussion of the method we refer to Ref.~\cite{Alarcon:2017ivh}.
\subsection{Leading-order calculation}
%
% FIGURE
% 
\begin{figure}
\begin{center}
\includegraphics[width=.36\textwidth]{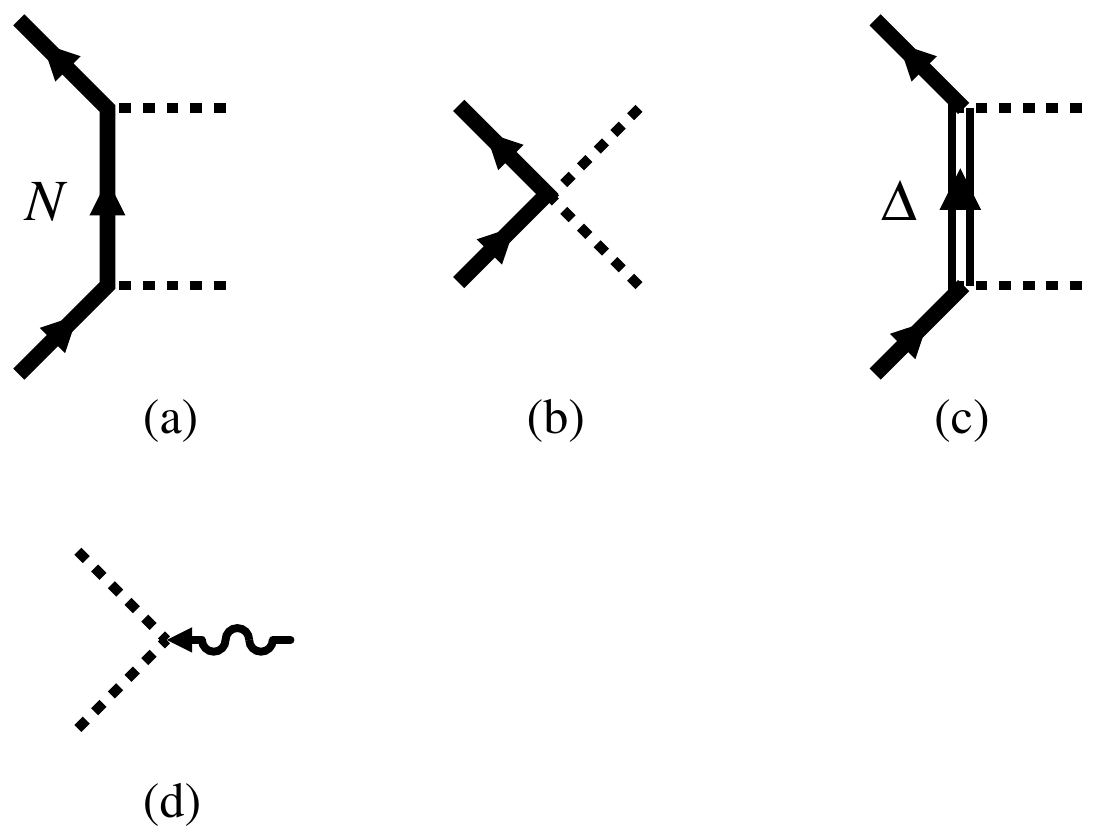}
\caption{\small (a, b, c)~LO $\chi$EFT diagrams contributing to the $\pi\pi \rightarrow N\bar N$
PWA in the $I = J = 1$ channel. (a)~$N$ Born term. (b)~Weinberg-Tomozawa contact term.
(c)~$\Delta$ Born term. (d)~Pion EM FF in LO.
\label{fig:eft}}
\end{center}
\end{figure} 
For calculating the $J$ functions of Eq.~(\ref{J_pm_def}) we use $\chi$EFT with the 
$SU(2)$-flavor group and relativistic $N$ and $\Delta$ degrees of freedom. 
This formulation ensures the correct position of the singularities and includes
the important contributions from the $\Delta$ Born term.
The setup of the $\chi$EFT calculation is described in Ref.~\cite{Alarcon:2012kn} and summarized 
in Ref.~\cite{Alarcon:2017asr} (fields, chiral Lagrangian, power counting, values of couplings). 
The interactions of the spin-3/2 $\Delta$ field are formulated with consistent 
vertices \cite{Pascalutsa:1998pw,Pascalutsa:1999zz, Pascalutsa:2000kd,Krebs:2008zb}, 
and the extended-on-mass-shell (EOMS) scheme is used to maintain the standard power counting
\cite{Fuchs:2003qc} (diagrams with pion loops do not enter in the present calculation).

The LO diagrams contributing to the $\pi\pi \rightarrow N \bar N$ PWAs
in the $I = J = 1$ channel are shown in Fig.~\ref{fig:eft}. They include the $N$ and $\Delta$ 
Born terms, Fig.~\ref{fig:eft}a and c, and the Weinberg-Tomozawa contact term, Fig.~\ref{fig:eft}b, 
which appears as the result of chiral invariance of the dynamics with relativistic baryons.
We take the results for the LO $\pi N \rightarrow \pi N$ amplitude of Ref.~\cite{Alarcon:2012kn} 
(the first relativistic $\chi$EFT calculation of $\pi N$ scattering with explicit $\Delta$), 
and project onto the $I = J = 1$ channel to get the PWAs $f_\pm^1 (t)$.
The pion EM FF at this order is just $F_\pi = 1$ (pointlike), see Fig.~\ref{fig:eft}d.
The $\chi$EFT results for $J_\pm^1(t)$ therefore coincide with $f_\pm^1(t)$ at this order.
Analytic expressions for $J_\pm^1(t)$ are given in Appendix~\ref{app:expressions}.
Numerical results of the LO approximation will be shown below.
\subsection{Estimates of higher-order corrections}
\label{subsec:higher_order}
%
% FIGURE
% 
\begin{figure}[t]
\begin{center}
\includegraphics[width=.36\textwidth]{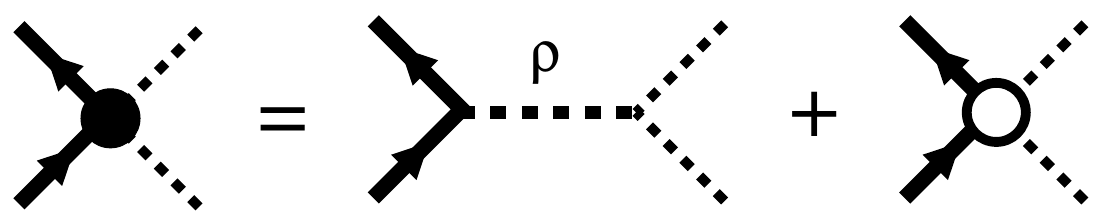}
\caption{\small Adjustment of the LEC of the NLO $\pi\pi NN$ contact term, $c_4$, in the DI$\chi$EFT approach. 
The original contact term (filled circle, left-hand side) is equated with the sum of $\rho$ meson 
exchange and the adjusted contact term (open circle).
\label{fig:contact}}
\end{center}
\end{figure} 
At NLO accuracy, corrections to the $I = J = 1$ $\pi\pi \rightarrow N\bar N$ PWAs
arise from the NLO contact term in the chiral Lagrangian with the low-energy constant (LEC) $c_4$. 
The value of this 
LEC has to be adjusted consistently with our unitarity-based approach \cite{Alarcon:2017asr}.
In standard $\chi$EFT calculations $c_4$ receives large contribution from $\rho$ meson exchange.
Since in our formulation the effect of the $\rho$ is included explicitly through $|F_\pi(t)|^2$, 
we have to remove it from the value of $c_4$ to avoid double-counting (see Fig.~\ref{fig:contact}).
Using the estimate for the $\rho$ contribution of Ref.~\cite{Bernard:1996gq},
$c_4^\rho \approx 1.63 \, \textrm{GeV}^{-1}$, and subtracting it from the empirical $c_4$ 
reported in Refs.~\cite{Alarcon:2012kn,Siemens:2016jwj}, we obtain the range
\be
c_4 \, [\textrm{adjusted}] \; = \; (-0.54, 0.27) \; \textrm{GeV}^{-1}.
\label{c4_adjusted}
\ee  
Note that these values are much smaller than the original $c_4$ and consistent with zero,
which means that the NLO corrections to the isovector spectral functions in our
formulation are very small. The analytic expressions for the NLO corrections to 
$J_\pm^1(t)$ are given in Appendix~\ref{app:expressions}.

At N2LO accuracy pion loop corrections appear, and the structure of the $\chi$EFT expressions 
becomes considerably richer. The $\pi\pi \rightarrow N\bar N$ PWAs and the pion FF now involve 
$\pi\pi$ rescattering in the $t$-channel and become complex at $t > 4 M_\pi^2$, 
in such a way that their phases cancel and the functions $J_\pm^1(t)$ of Eq.~(\ref{J_pm_def})
remain real. Also, $\pi N$ and $\pi\Delta$ $s$-channel intermediate states appear in the 
$\pi N$ amplitude.
Following Ref.~\cite{Alarcon:2017asr} we perform a simple estimate of the N2LO corrections 
to the spectral functions of the electric FF, by taking the N2LO tree-level amplitudes and fixing 
the LECs through the charge sum rule. We require that the unsubtracted dispersion relation 
for the isovector electric FF reproduce the isovector charge when the integration is restricted 
to the region $t' < t_{\rm max} \sim 1\, \textrm{GeV}^2$,
\be
\frac{1}{\pi}
\int_{4 M_\pi^2}^{t_{\rm max}} dt' \; \frac{\textrm{Im}\, G_E^V (t')}{t'} 
\;\; = \;\; {\textstyle\frac{1}{2}}.
\ee
This condition gives N2LO contact term contributions with sign opposite to that of the 
LO and NLO results, which provides a crucial curvature in the electric spectral function 
and allows us to extend the calculations up to $t\sim 1\, \textrm{GeV}^2$.
In the language of traditional dispersion analysis these contact terms represent the
negative contributions from the $\rho'$, which compensate the excess charge
that would be produced by the $\rho$ alone. 
In the magnetic FF no new tree-level amplitudes with LECs enter at N2LO level,
so that the described method of estimating of the corrections cannot be applied.
Our calculations of $G_M^V$ are therefore limited to NLO accuracy, and are expected
to describe the empirical spectral functions only at $t\ll 1\, \textrm{GeV}^2$.
Numerical results of the NLO and N2LO approximations will be shown below.
\subsection{Timelike pion EM form factor}
\label{subsec:pion}
%
% FIGURE
% 
\begin{figure}
\begin{center}
\includegraphics[width=.48\textwidth]{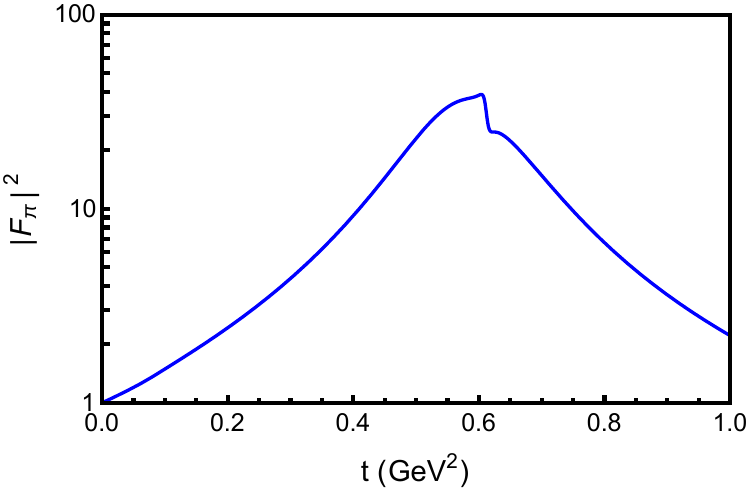}
\caption{\small Empirical parametrization of the timelike pion EM FF $|F_\pi(t)|^2$
obtained from $e^+e^- \rightarrow \pi^+\pi^-$ annihilation data (details see text).
\label{fig:pion}}
\end{center}
\end{figure} 
For evaluating the timelike pion FF entering in our calculation we use a Gounaris-Sakurai 
parametrization of the $e^+e^- \rightarrow \pi^+\pi^-$ exclusive annihilation data \cite{Gounaris:1968mw},
including effects of $\rho$-$\omega$ mixing \cite{Barkov:1985ac}, with the parameters determined 
in Ref.~\cite{Lorenz:2012tm}. The squared modulus $|F_\pi(t)|^2$ is shown in Fig.~\ref{fig:pion}.
One clearly sees the $\rho$ resonance at $t \sim 0.6\, \textrm{GeV}^2$ and the rapid
variation resulting from $\rho$--$\omega$ mixing. The fact that $|F_\pi(t)|^2$
reaches a value of $\sim$2 at $t \sim 0.2\, \textrm{GeV}^2$, and $\sim 10$ at 
$t \sim 0.4\, \textrm{GeV}^2$, shows that $\pi\pi$ rescattering is very substantial already
at moderate $t$ and justifies our approach of incorporating these effects empirically.

Since $|F_\pi(t)|^2$ is determined very accurately from the annihilation data we neglect the effect
of its uncertainty on the spectral functions. In the following we quote only the uncertainties of 
the spectral function resulting from the $\chi$EFT calculation of $J_\pm^1(t)$.
\section{Results}
\label{sec:results}
\subsection{Isovector EM spectral functions}
%
% FIGURE
%
\begin{figure}
\begin{center}
\includegraphics[width=.45\textwidth]{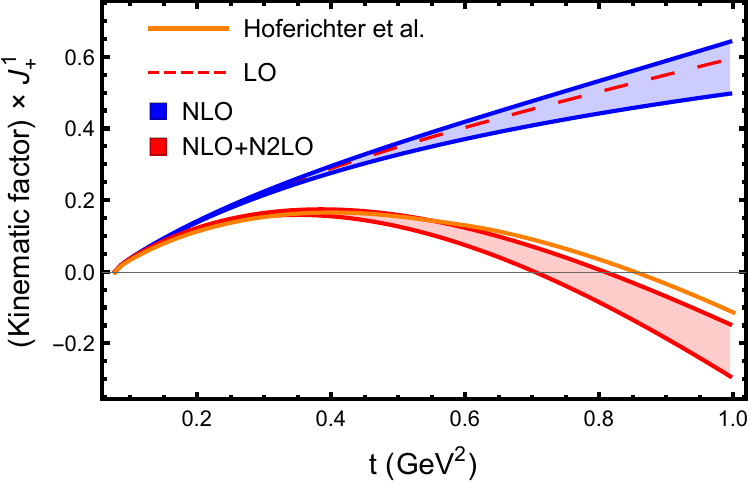} \\
\includegraphics[width=.45\textwidth]{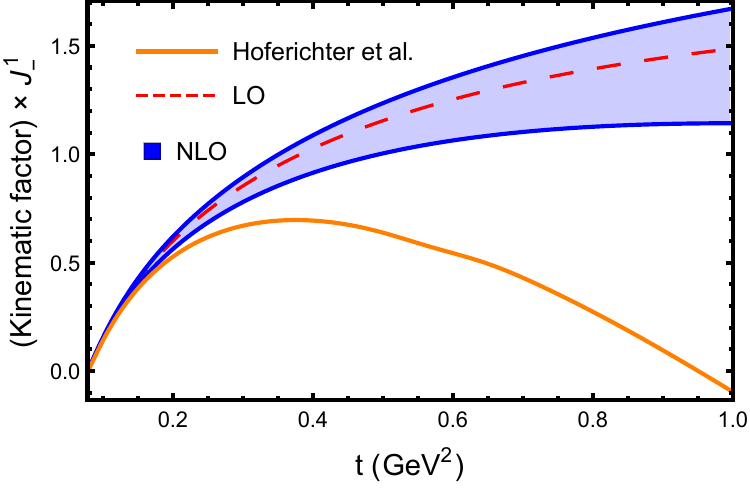}
\caption{\small $\chi$EFT results for the functions 
$k_{\rm cm}^3/(m_N \sqrt{t}) \, J^1_+(t)$ and $k_{\rm cm}^3/(\sqrt{2t}) \, J^1_-(t)$, 
Eqs.(\ref{unitarity_G_E_real})--(\ref{J_pm_def}), at $t > 4 M_\pi^2$.
Dashed lines: LO approximation. Blue bands: NLO approximation. Red band: NLO+N2LO, estimated 
as described in Sec.~\ref{subsec:higher_order}. 
Solid orange lines: Roy-Steiner analysis results \cite{Hoferichter:2016duk}.
\label{fig:Js}.}
\end{center}
\end{figure} 
The spectral functions of the isovector EM nucleon FFs are evaluated using
the DI$\chi$EFT method and parameters described in Sec.~\ref{sec:calculation}.
Figure~\ref{fig:Js} shows the results of the $\chi$EFT calculation of the real 
functions $J^1_\pm (t)$ at $t > 4 M_\pi^2$, Eq.~(\ref{J_pm_def}). For a better view the plot shows 
the functions multiplied by the kinematic factors of 
Eqs.~(\ref{unitarity_G_E_real}) and (\ref{unitarity_G_M_real}), 
$k_{\rm cm}^3/(m_N \sqrt{t})$ and $k_{\rm cm}^3/\sqrt{2t}$, respectively.
One observes: (a)~The $J^1_\pm (t)$ with the kinematic factors are smooth functions, 
as expected on grounds of their analytic properties. (b)~The $\chi$EFT 
calculations of $J^1_\pm (t)$ show good convergence. In both functions higher-order corrections 
are small at threshold and increase with $t$. LO and NLO results are close because the 
adjusted LEC $c_4$ is small, Eq.~(\ref{c4_adjusted}). In $J^1_+ (t)$ the N2LO 
corrections, estimated as described in Sec.~\ref{subsec:higher_order}, are
negative and cause the function to decrease and turn negative at larger $t$.
(c)~The $\chi$EFT results for  $J^1_\pm (t)$ show reasonable agreement with the functions extracted
from an analysis of $\pi N$ scattering data using Roy-Steiner equations \cite{Hoferichter:2016duk}.
In both $J^1_+$ and $J^1_-$ the LO and NLO approximation agree with the Roy-Steiner
result up to $t \sim 0.2\, \textrm{GeV}^2$. In $J^1_+$ the negative N2LO corrections
extend the region of agreement up to $t \sim 1\, \textrm{GeV}^2$.

\begin{figure}
\begin{center}
\epsfig{file=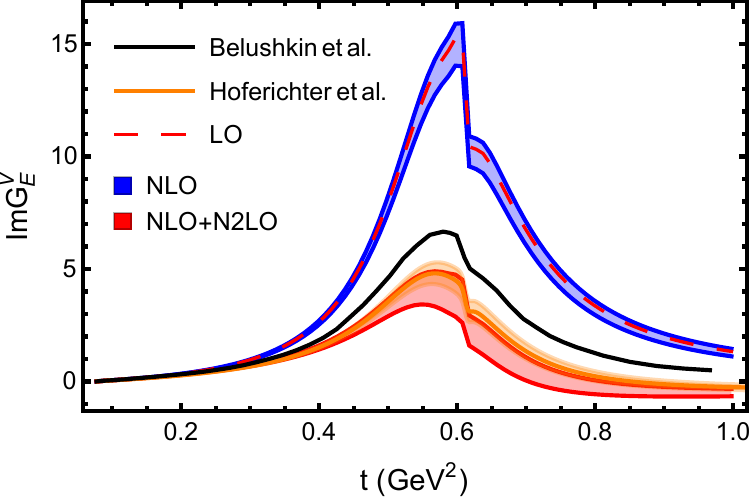,width=.45\textwidth,angle=0}
\epsfig{file=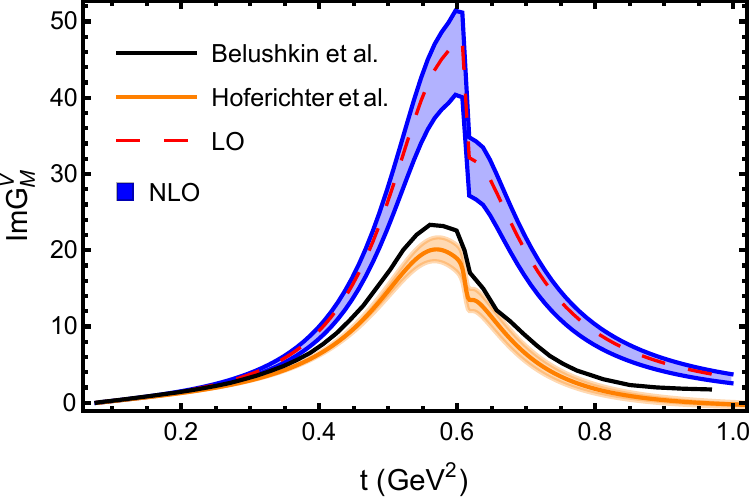,width=.45\textwidth,angle=0}
\caption{\small 
DI$\chi$EFT results for the spectral functions of the isovector nucleon EM FFs,
Eqs.(\ref{unitarity_G_E_real}) and (\ref{unitarity_G_M_real}).
Dashed line: LO approximation. Blue band: NLO approximation. Red band: NLO+N2LO, estimated
as described in in Sec.~\ref{subsec:higher_order}). 
Orange band: Results of Roy-Steiner analysis \cite{Hoferichter:2016duk}.
Black line: Empirical dispersion analysis \cite{Belushkin:2005ds}. 
\label{Fig:ImGEGM}}
\end{center}
\end{figure}

Figure~\ref{Fig:ImGEGM} shows the isovector spectral functions, obtained by 
multiplying the $\chi$EFT results for $J^1_\pm (t)$ with the empirical $|F_\pi(t)|^2$
cf.\ Eqs.~(\ref{unitarity_G_E_real}) and (\ref{unitarity_G_M_real}).
The spectral functions clearly show the effects of $\pi\pi$ rescattering, which are
not suitable for perturbative $\chi$EFT treatment and included through the 
empirical pion FF in our approach. Note that the enhancement through $|F_\pi(t)|^2$ 
is large even near the two-pion threshold $t \sim 4\, M_\pi^2$, cf.\ Sec.~\ref{subsec:pion} and
Fig.~\ref{fig:pion}. The convergence pattern of the spectral functions
follows from that of the $\chi$EFT calculation of $J_\pm^1(t)$. 
In both $\textrm{Im} \, G_E$ and $\textrm{Im} \, G_M$, the LO and NLO approximations
are in good agreement with the Roy-Steiner results up to $t \sim 0.2\, \textrm{GeV}^2$.
In $\textrm{Im} \, G_E$ the negative N2LO correction (estimated) is sufficient to reproduce 
the Roy-Steiner result up to $t \sim 1\, \textrm{GeV}^2$.
%
% FIGURE
%
\subsection{Nucleon EM radii}
\begin{figure}
\begin{center}
\includegraphics[width=.48\textwidth]{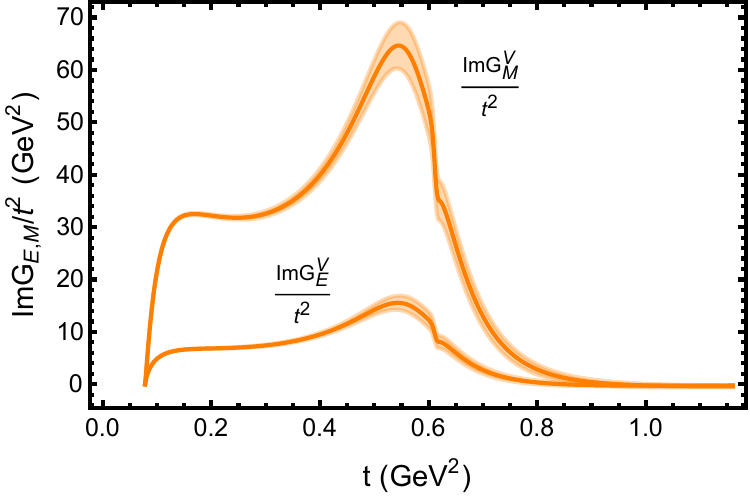}
\caption{\small The integrands of the dispersive integrals for the isovector nucleon
electric and magnetic radii, Eq.~(\ref{r2_dispersion}),
evaluated with the spectral functions of Ref.~\cite{Hoferichter:2016duk}.
\label{fig:disp_r2}}
\end{center}
\end{figure}

The nucleon's isovector electric and magnetic radii are given by the dispersion integrals
\be
\langle r^2 \rangle_i^V
\;\;  = \;\; \frac{6}{\pi} \int_{4M_\pi^2}^\infty dt' \; \frac{\text{Im}\, G_i^V (t')}{t'^{2}}
\hspace{2em} (i = E, M).
\label{r2_dispersion}
\ee
The factor $1/{t'}^2$ ensures convergence of the integral over the range $t' \lesssim 1\, \textrm{GeV}^2$;
see Fig.~\ref{fig:disp_r2}. The integrals can therefore be evaluated with the DI$\chi$EFT spectral functions. 
Table~\ref{Tab:Isovector_radii_GEMV} summarizes the DI$\chi$EFT results for the isovector radii and
compares them with the results of other dispersive approaches and LQCD calculations.
For the electric radius our NLO+N2LO result agrees very well with the result of the
Roy-Steiner analysis \cite{Hoferichter:2016duk}. Our lower-order results overestimate this value, 
because the LO and NLO spectral functions are larger than the Roy-Steiner result
around the $\rho$ peak. For the magnetic radius our result is larger by a factor $\sim$2 
than the phenomenological and LQCD results, because our magnetic spectral function is
likewise too large around the $\rho$ peak.
%
% TABLE
%
\begin{table*}
\begin{ruledtabular}
\begin{tabular}{cccc|ccccc}
  &LO  & NLO & NLO+N2LO & Lorenz & Epstein &
  Hoferichter & LQCD & Leupold \\[-.6ex] &&&&
  \cite{Lorenz:2012tm} &\cite{Epstein:2014zua}
  &\cite{Hoferichter:2016duk} &
  \cite{Alexandrou:2017ypw}&\cite{Leupold:2017ngs} \\[1ex]
$\langle r^2 \rangle_E^V$ (fm$^2$)& 0.98 & (0.98 , 0.99) & (0.33 ,
  0.43) & 0.416(8) & -- &0.405(36) & 0.327(24)(15) & (0.27,0.31) \\ $\langle
  r^2 \rangle_M^V$ (fm$^2$)& 3.28 & (2.87 , 3.50) & -- &
  $1.78^{+0.10}_{-0.11} $ & 1.81(7) & 1.81(11) & 1.08(11)(14) &
  1.81\textsuperscript{a} \\
\end{tabular}
\caption{Isovector nucleon EM radii calculated in DI$\chi$EFT (left columns)
and in other approaches (right columns). \\
\textsuperscript{a}In the $\chi$EFT calculation of Ref.~\cite{Leupold:2017ngs} the LEC $c_4$
is adjusted to reproduce the magnetic radius.
\label{Tab:Isovector_radii_GEMV}}
\end{ruledtabular}
\end{table*}
%
% FIGURE
%

DI$\chi$EFT allows us to calculate the isovector nucleon EM FFs and radii, 
which are matrix elements of $G$--parity even operators. Experiments measure the
individual proton and neutron FFs and radii. In view of the questions concerning
the proton charge radius measurements it is interesting to compare our results 
directly with the experimental results for the proton and neutron charge radii.
To do so, we supplement the DI$\chi$EFT results for the isovector spectral functions
with an empirical parametrization of the isoscalar spectral functions.
We use a two-pole parametrization with the $\omega$ pole at $M_\omega^2 = 0.61 \, \textrm{GeV}^2$
and a second pole at $M_2^2 \approx M_\phi^2 = 1 \, \textrm{GeV}^2$
\cite{Alarcon:2017asr} 
\ba
\textrm{Im} \, G_i^S(t) &=& \pi \left[ a_i^\omega \delta (t - M_\omega^2)
+ a_i^{(2)} \delta (t - M_2^2) \right] 
%\textrm{Im} \, G_i^S(t) \; = \; \pi \sum_{V = \omega, \phi} a_i^V \delta (t - M_V^2)
\nonumber \\
&&
\hspace{1em} (i = E, M),
\label{isoscalar_empirical}
\ea
where the coefficients $a_i^\omega$ (including their uncertainties) are taken from the 
dispersive FF fit of Ref.~\cite{Belushkin:2006qa}, and the coefficients $a_i^{(2)}$ 
are adjusted to reproduce the total charge and magnetic moment. The second pole 
is an effective pole representing the overall strength of the spectral 
function at $t \sim 1\, \textrm{GeV}^2$; the details of the strength distribution at these
values of $t$ are not important for estimating the nucleon radii.
The proton and neutron radii obtained in this way are summarized in Table~\ref{Tab:Isovector_radii_GEM}.
In the proton and neutron electric radii, the estimated uncertainty is dominated by the isoscalar 
component, which is purely empirical. Our results obtained with the
NLO+N2LO DI$\chi$EFT calculation of the isovector radii are in agreement with the experimental values.
In the magnetic radii the estimated uncertainty is likewise dominated by the isoscalar component.
Note that the DI$\chi$EFT calculation of the isovector magnetic spectral function does not include 
the N2LO corrections and strongly overestimates the empirical result 
(cf.\ Table~\ref{Tab:Isovector_radii_GEMV});
this discrepancy is not reflected in the uncertainty estimate.
\begin{table*}
\begin{ruledtabular}
\begin{tabular}{cccc|cc}
 &LO & NLO & NLO+N2LO & LQCD \cite{Alexandrou:2017ypw}& PDG \cite{Patrignani:2016xqp} \\
  $\langle r^2 \rangle_E^p$ (fm$^2$)& (1.11, 1.49) & (1.05, 1.52) & (0.46, 0.94  ) 
& 0.589(39)(33)  & (0.706, 0.707) \\[-.6ex]
& & & &          & (0.755, 0.777) \\[.6ex]
  $\langle r^2 \rangle_M^p$ (fm$^2$)& (1.19, 1.46) & (1.04, 1.54) & -- & 0.506(51)(42) & (0.53 , 0.68)  \\
  $\langle r^2 \rangle_E^n$ (fm$^2$)& $(-0.84, -0.47)$ & $(-0.88, -0.40)$ & $(-0.29, 0.18)$ & 
    $-0.038(34)(6)$   &  $-0.1161(22)$  \\
  $\langle r^2 \rangle_M^n$ (fm$^2$)& (1.29, 1.64) & (1.08, 1.81) & -- & 0.586(58)(75) & (0.73, 0.76)  \\
\end{tabular}
\caption{Left columns: Proton and neutron EM radii obtained from the DI$\chi$EFT calculation of the isovector 
radii and the empirical parametrization of the isoscalar radii. Right columns: LQCD and experimental
results. For the experimental values we quote the averages compiled by the Particle Data Group (upper and 
lower limits obtained by adding statistical and systematic errors) \cite{Patrignani:2016xqp}. 
The smaller values correspond to the extraction from muonic hydrogen measurements; the larger values
correspond to electronic hydrogen measurements; extractions from electron-proton scattering data
using different methods have produced results supporting either value; see Ref.~\cite{Patrignani:2016xqp} 
for details.
\label{Tab:Isovector_radii_GEM}}
\end{ruledtabular}
\end{table*}
\subsection{Higher derivatives of EM form factors}
Higher derivatives of the nucleon EM FFs at $t = 0$ are of interest for several reasons. In the 
experimental analysis, the values of higher derivatives allowed (or assumed) in fits of FF data 
at $t < 0$ directly affect the extrapolation to $t = 0$ and extraction of the nucleon charge radii; 
see Refs.~\cite{Lee:2015jqa,Higinbotham:2015rja,Griffioen:2015hta,Horbatsch:2016ilr,Sick:2017aor}
for details. In the theoretical studies reported here, higher derivatives of the FFs represent 
clean chiral observables that can be predicted almost model-independently with minimal uncertainties.
The comparison of low- and high-order derivatives reveals the presence of two dynamical scales 
in the nucleon FFs, which implies a surprisingly rich structure and should be incorporated
into the experimental analysis.

In the context of the traditional representation of the FFs as Fourier transforms of 
3-dimensional spatial densities, the higher derivatives of the FFs at $t = 0$ correspond 
to the higher $r^2$--weighted moments of the densities. The connection is given by \cite{Horbatsch:2016ilr}
\ba
G_E(t) &=& 1 + \frac{\langle r^2 \rangle_E}{3!}t + \frac{\langle r^4 \rangle_E}{5!}t^2 
+ \frac{\langle r^6 \rangle_E}{7!}t^3 + \ldots, 
\hspace{1em}
\label{higher_moments_E} \\
\frac{G_M(t)}{\mu} &=& 1 + \frac{\langle r^2 \rangle_M}{3!}t + \frac{\langle r^4 \rangle_M}{5!}t^2 
+ \frac{\langle r^6 \rangle_M}{7!}t^3 + \ldots,
\hspace{1em}
\label{higher_moments_M}
\ea
\vspace{-1ex}
\ba
\frac{1}{n!} \frac{d^n G_E}{dt^n} (0) &=& \frac{\langle r^{2n} \rangle_E}{(2n+1)!} ,
\label{derivative_moment_E}
\\[1ex] 
\frac{1}{n! \, \mu} \frac{d^n G_M}{dt^n} (0) &=& \frac{\langle r^{2n} \rangle_M}{(2n+1)!}
\label{derivative_moment_M}
\\[2ex] \nonumber
&& \textrm{(for either $p$ or $n$).}
\ea
Note that for the proton and neutron the magnetic radii are defined as the derivatives of the 
FFs divided by the magnetic moments; this is not the case for the isovector and isoscalar components. 
In the following we quote results for the moments; they can be converted to FF derivatives
through Eqs.~(\ref{derivative_moment_E}) and (\ref{derivative_moment_M}).\footnote{The representation
of FFs in terms of 3-dimensional spatial densities is physically meaningful only for nonrelativistic systems.
For relativistic systems such as the nucleon a proper spatial representation is provided by
the 2-dimensional transverse densities at fixed light-front time; see Ref.~\cite{Miller:2010nz}
for a review. We refer to the moments $\langle r^{2n} \rangle$ only because this representation 
is used in the experimental literature, and use it only in the sense of a mathematical representation 
of the FF derivatives at $t = 0$.}

%
% FIGURE
%
\begin{figure}
\begin{center}
\includegraphics[width=.48\textwidth]{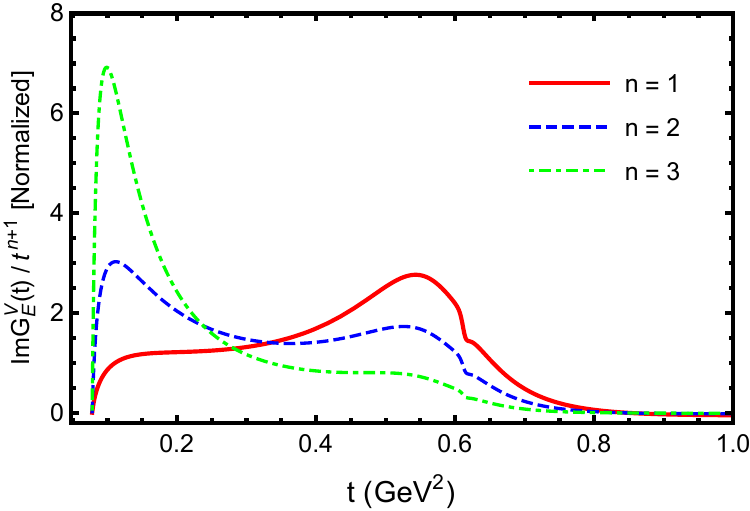}
\caption{\small Integrand of the dispersive integral for the moments of the 
isovector electric FF $G_E^V$, Eq.~(\ref{Eq:Derivatives_FF}), for $n = 1, 2$ and $3$,
evaluated with the spectral functions of Ref.~\cite{Hoferichter:2016duk}. The plot shows
the $t'$ distributions divided by the value of the integral, i.e., normalized to unit 
area under the curves.
\label{fig:disp_r2_higher}}
\end{center}
\end{figure} 
The higher moments of the isovector FFs are given by the dispersion integrals
\be
%\left.\frac{d^nG_i^V(t)}{dt^n}\right|_{t=0} 
\frac{\langle r^{2n} \rangle_i^V}{(2n+1)!}
\; = \; \frac{1}{\pi} \int_{4M_\pi^2}^\infty dt' \; \frac{\text{Im} \, G_i^V(t')}{t'^{n+1}}
\hspace{1em} (i = E, M).
\label{Eq:Derivatives_FF}
\ee
The factors $1/t'^{n + 1}$ strongly suppress contributions from large $t'$ and render
the integrals well convergent. The integrals can therefore be evaluated accurately using 
the DI$\chi$EFT spectral functions. Figure~\ref{fig:disp_r2_higher} shows the integrands
for the isovector electric FF derivatives with $n = \{1, 2, 3\}$; the curves are normalized 
to unit integral for each $n$ and show the relative distribution of strength in $t'$.
The distributions clearly show the presence of two dynamical scales: the $\rho$ meson mass 
$m_\rho^2 \sim 0.6\, \textrm{GeV}^2 \approx 30\, M_\pi^2$ (the peak of the spectral function), 
and the two-pion threshold $4 M_\pi^2$ (the start of the spectral integral). The integrals 
receive contributions from both regions of $t'$, and their relative importance changes with $n$.
For a rough assessment we can take $M_\rho^2/2 = 0.3\, \textrm{GeV}^2$ as the boundary 
between the two regions. For $n = 1$ approximately 2/3 of the integral comes from the region 
$t' > M_\rho^2/2$, and $1/3$ from $4 M_\pi^2 < t < M_\rho^2/2$. For $n = 2$, each region 
contributes about 1/2. For $n = 3$ and higher, the near-threshold region dominates.

%
% FIGURE
%
\begin{figure}
\begin{center}
\includegraphics[width=.48\textwidth]{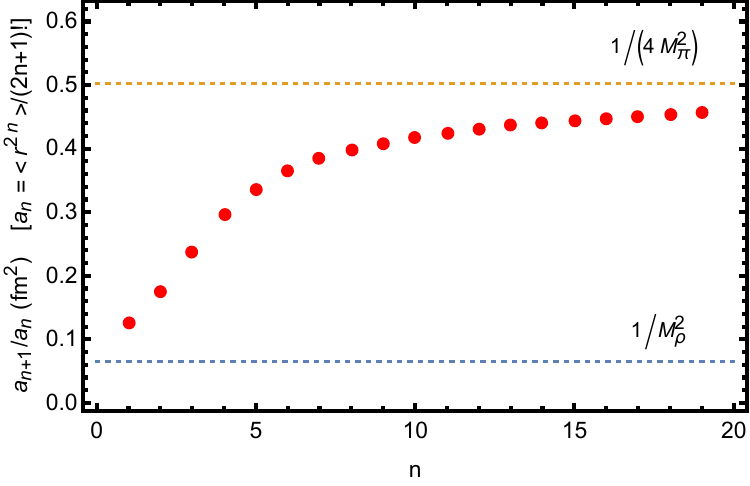}
\caption{\small Ratios of successive moments of the isovector electric FF $G_E^V$,
Eq.~(\ref{moment_ratio}), computed using the DI$\chi$EFT isovector spectral functions (see Fig.~\ref{Fig:ImGEGM}).
The horizontal lines indicate the values of the dynamical scales $1/M_\rho^2$ and $1/(4 M_\pi^2)$
\label{fig:r2_e_ratio}}
\end{center}
\end{figure} 
The presence of two dynamical scales in the isovector moments can also be demonstrated by considering
the ratios of successive moments,
\be
\left. \frac{\langle r^{2n + 2} \rangle_i^V}{(2n+3)!} \; \right/
\frac{\langle r^{2n} \rangle_i^V}{(2n+1)!}
\hspace{2em} (i = E, M).
\label{moment_ratio}
\ee
If the dispersion integral were dominated by a certain region of $t'$, the value of the
ratio Eq.~(\ref{moment_ratio}) would be given by the average of $1/t'$ over that region.
The ratios thus directly reveal the effective values of $1/t'$ in the integral.
Figure~\ref{fig:disp_r2_higher} shows the ratios of the isovector electric FF moments
obtained with the DI$\chi$EFT spectral functions. One sees that the ratios start with
a value $\sim 1/m_\rho^2$ at $n = 1$ and increase to values $\sim 1/(4 M_\pi^2)$ at large $n$.

The presence of two dynamical scales implies that the higher FF moments are of ``unnatural'' size, 
i.e., their values are very different from what one would estimate using the value of the
lowest moment and a single-scale functional form of the FF. [In the dispersive representation such a 
single-scale form would be e.g.\ a spectral function $\textrm{Im}\, G_{i}^V \propto \delta (t - M_\rho^2)$, 
or derivatives thereof.] This conclusion relies only on general features of the dispersive
representation and is insensitive to the details of the dynamical calculation presented here.
It has has numerous consequences for the interpretation of the FF moments and the analysis of 
low-$t$ elastic scattering experiments, which will be elaborated below.

%
% TABLE
%
\begin{table}
\begin{ruledtabular}
\begin{tabular}{lccc}
$G_E^V$ &  &  & \\
\hline
&LO & NLO &  NLO+N2LO\\
    $ \langle r^4 \rangle $ (fm$^4$)       &   1.81 & (1.72, 1.86)  & (0.88, 1.02) \\
    $ \langle r^6 \rangle $ (fm$^6$)       &   9.86 & (9.54, 10.03) & (6.68, 7.16) \\ 
    $ \langle r^8 \rangle $ ($10^2$ fm$^8$)&   1.40 & (1.37, 1.41)  & (1.17, 1.20) \\ 
\hline
$G_M^V$ &  &  & \\
\hline
&LO & NLO & \\
    $ \langle r^4 \rangle $ (fm$^4$)        &  6.49 & (5.81, 6.85)  & \\
    $ \langle r^6 \rangle $ (10 fm$^6$)     &  3.82 & (3.53, 3.98)  & \\ 
    $ \langle r^8 \rangle $ ($10^2$ fm$^8$) &  5.68 & (5.38, 5.84)  & \\ 
\end{tabular}
\end{ruledtabular}
\caption{Higher moments of the isovector nucleon FFs calculated in DI$\chi$EFT.}
\label{tab:higher_derivatives_isovector}
\end{table}
Table~\ref{tab:higher_derivatives_isovector} shows the DI$\chi$EFT results for the
higher moments of the isovector FFs, $G_E^V$ and $G_M^V$. Because the dispersion
integrals with $n \geq 2$ sample the spectral functions near threshold, the higher moments 
can be computed accurately and represent genuine predictions of our approach. 
This is seen in the intrinsic uncertainty estimates of Table~\ref{tab:higher_derivatives_isovector}:
with increasing $n$, the derivatives become less sensitive to higher-order chiral corrections.
We emphasize that one should be careful in interpreting the numerical
values of the individual moments in Table~\ref{tab:higher_derivatives_isovector},
as they contain large factorial factors. The unnatural behavior of the higher moments should be
demonstrated by taking ratios (see above) or comparing the moments to a reference FF (see below).

It is worth noting that the $\pi\pi$ rescattering effects included through the timelike pion FF play an 
important role even in the higher FF moments. The moments with $n \geq 3$ receive most of their contributions
from the region $4 M_\pi^2 < t' \lesssim 10 \, M_\pi^2$ of the spectral integral Eq.~(\ref{Eq:Derivatives_FF})
(see Fig.~\ref{fig:disp_r2_higher}). The value of $|F_\pi(t')|^2$ is $\sim 1.3$ at $t' = 4 M_\pi^2$, 
and $\sim 2$ at $10 M_\pi^2$ (see Fig.~\ref{fig:pion}). The enhancement compared to traditional
direct $\chi$EFT calculations of the spectral functions is therefore quite substantial (note that without
the factor $|F_\pi|^2$ our results at LO and NLO would be identical to those of the direct calculation).
Nevertheless our results for the higher moments agree with those of the direct $\chi$EFT calculation 
of Ref.~\cite{Peset:2014jxa} within errors.

%
% TABLE
%
\begin{table}
\begin{ruledtabular}
\begin{tabular}{lcccc}
$G_E^p$ &  &  & \\
\hline
 &LO & NLO &  NLO+N2LO\\
  $\langle r^4 \rangle$ (fm$^4$)       & (2.09, 2.48)  & (2.00, 2.53) & (1.16, 1.70) \\
  $\langle r^6 \rangle$ (fm$^6$)       & (10.8, 11.7)  & (10.5, 11.9) & (7.59, 9.00) \\
  $\langle r^8 \rangle$ ($10^2$ fm$^8$)& (1.44, 1.48)  & (1.42, 1.49) & (12.1, 12.9) \\
  \hline
$G_E^n$ &  & & \\

\hline  &LO & NLO  &  NLO+N2LO\\
  $\langle r^4 \rangle$ (fm$^4$)      & (-1.53, -1.13) \! & (-1.58, -1.04) & (-0.74, -0.20) \\
  $\langle r^6 \rangle$ (fm$^6$)      & (-8.94, -8.02) \! & (-9.11, -7.71) & (-6.24, -4.84) \\
  $\langle r^8 \rangle$ ($10^2$ fm$^8$)& (-1.35, -1.31) \! & (-1.36, -1.29) & (-1.15, -1.08) \\
  \hline
$G_M^p$ &  & & \\
\hline  &LO & NLO &\\
  $ \langle r^4 \rangle$ (fm$^4$)       & (2.38, 2.68) & (2.14, 2.81) \\
  $ \langle r^6 \rangle$ (10 fm$^6$)    & (1.39, 1.46) & (1.29, 1.52) \\
  $ \langle r^8 \rangle$ ($10^2$ fm$^8$)& (2.05, 2.08) & (1.94, 2.13) \\
  \hline
$G_M^n$ &  & & \\
\hline  &LO & NLO & \\
  $ \langle r^4 \rangle$ (fm$^4$)       &  (3.30, 2.87) & (3.49, 2.51) \\
  $ \langle r^6 \rangle$ (10 fm$^6$)    &  (1.96, 1.86) & (2.04, 1.71) \\
  $ \langle r^8 \rangle$ ($10^2$ fm$^8$)&  (2.95, 2.91) & (3.04, 2.75) \\
\end{tabular}
\caption{Higher moments of the proton and neutron electric and magnetic FFs,
calculated using the DI$\chi$EFT results for the isovector moments
and the empirical parametrization of the isoscalar FFs.
\label{Tab:Moments_2}}
\end{ruledtabular}
\end{table}
In the isoscalar FFs the strength of the spectral functions is located overwhelmingly at 
the $\omega$ meson mass. The higher moments are therefore governed by this single scale and 
are of natural size. This in turn implies that the higher moments of the proton and neutron 
FFs are dominated by the isovector component and can be inferred from our DI$\chi$EFT results.
Table~\ref{Tab:Moments_2} shows our results for the moments of $G_E^p, G_E^n, G_M^p$ and
$G_M^n$, obtained using the DI$\chi$EFT results for the isovector moments and the 
empirical parametrization of the isoscalar FFs. We stress that the isoscalar information 
is used here only to demonstrate that the higher derivatives are dominated by the 
isovector component, and that the uncertainties associated with the isoscalar parametrization 
are irrelevant in the higher derivatives.

\begin{figure}
\begin{center}
\includegraphics[width=.48\textwidth]{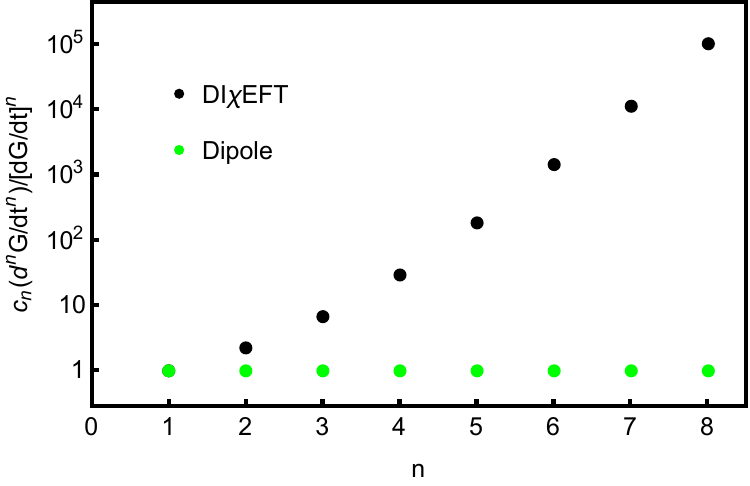}
\caption{\small The normalized ratios of the $n$th derivative of the proton electric FF $G_E^p$
and the $n$th power of the first derivative, Eq.~(\ref{ratio_dipole}). The ratios are normalized 
such that their values are unity for the dipole FF Eq.~(\ref{dipole}).
\label{fig:dipole}}
\end{center}
\vspace{-1ex}
\end{figure}

The theoretical results described here have implications for the analysis of electron-proton 
elastic scattering data at low $Q^2 \equiv -t$ and the extraction of the proton charge radius. 
The overall behavior of $G_E^p$ in the region $0 < Q^2 \lesssim 1\, \textrm{GeV}^2$ is associated
with a scale of the order of the vector meson mass $M_V^2 \; (V = \rho, \omega)$. The first derivative 
of $G_E^p$ at $Q^2 = 0$ is of the order $1/M_V^2$ and therefore appears natural, i.e., simple single-scale 
parametrizations of the finite-$Q^2$ data give a reasonable estimate of the first derivative.
The higher derivatives, however, are governed by the scale $1/(4 M_\pi^2)$ and appear unnatural.
Single-scale parametrizations or ``natural'' powers of the first derivative the data give 
qualitatively wrong estimates of the higher derivatives. To illustrate the point we compare the 
order-of-magnitude of the higher derivatives obtained from DI$\chi$EFT with the ones of the 
dipole parametrization
\be
G_E^p(t)[\textrm{dipole}] = \Lambda^4 / (t - \Lambda^2)^2 ,
\label{dipole}
\ee
which provides a good overall description of the FF data at $0 < Q^2 \lesssim 1\, \textrm{GeV}^2$
with $\Lambda^2 \approx 0.71\, \textrm{GeV}^2$. Figure~\ref{fig:dipole} shows the ratios
\be
\left. c_n \frac{d^n G_E^p}{dt^n} (0) \right/ \left[ \frac{d^nG_E^p}{dt^n} (0) \right]^n,
\hspace{2em} c_n \equiv \frac{2^n}{(n + 1)!} ,
\label{ratio_dipole}
\ee
as obtained with the DI$\chi$EFT results. The coefficients $c_n$ are defined such that for the 
dipole FF Eq.~(\ref{dipole}) the ratio is equal to unity for all $n$. The ratio Eq.~(\ref{ratio_dipole}) 
therefore indicates how strongly the actual higher derivatives deviate from the single-scale estimate based
on the dipole form. One sees that the ratio is $\sim$10$^2$ for $n = 4$, and reaches values
$\sim$10$^5$ for $n = 8$. It shows the striking consequences of the two dynamical scales in the 
higher FF derivatives, as implied by their dispersive representation. A similar observation regarding
the ratio of the actual FF moments to the dipole was made in the context of an empirical analysis 
in Ref.~\cite{Distler:2010zq}.
%
% FIGURE
%
%

%
% TABLE
%
\begin{table*}
\begin{ruledtabular}
\begin{tabular}{ccccccc}
$G_E^p$ &  &  &  & & \\
\hline
& Standard & Bernauer &  Horbatsch 
& Higinbotham 
& Higinbotham 
& Sick  \\[-.6ex]
& Dipole & \cite{Bernauer:2010zga} & \cite{Horbatsch:2016ilr} & Rational \cite{Higinbotham:2015rja} 
& Power series \cite{Higinbotham:2015rja} &\cite{Sick:2017aor} \\[1ex]
  $ \langle r^4 \rangle$  (fm$^4$)  &  1.08 &   2.63 &  0.60 &  1.38 & 1.50  & 2.01(5) \\
  $ \langle r^6 \rangle$  (fm$^6$)  &  3.30 &  26.93 &  5.00 &  5.68 & 5.75  &               \\
  $ \langle r^8 \rangle$  (fm$^8$)  & 16.2  & 408.12 & 99.36 & 40.06 & 24.68 &               \\
\end{tabular}
\caption{Higher moments of $G_E^p$ extracted from recent fits to low-$Q^2$ 
FF data using different classes of functions.
\label{Tab:Moments_Comparison}}
\end{ruledtabular}
\end{table*}
The values of the higher derivatives of $G_E^p$ and their impact on the $Q^2 \rightarrow 0$ 
extrapolation are presently the subject of intense 
discussions \cite{Lee:2015jqa,Higinbotham:2015rja,Griffioen:2015hta,Horbatsch:2016ilr,Sick:2017aor}. 
Fits to the low-$Q^2$ FF data
with different classes of functions (polynomials, rational functions) give widely different 
values of the second and higher derivatives; see Table~\ref{Tab:Moments_Comparison}
for a compilation of recent results. The DI$\chi$EFT results are broadly consistent with the
range of empirical values. An analysis of FF data incorporating theoretical constraints from 
DI$\chi$EFT will be the subject of a future study \cite{AHW}. For reference we quote in 
Appendix~\ref{app:derivatives} the numerical values of the DI$\chi$EFT moments of $G_E^p$ up 
to $n = 20$. While the individual values have little physical significance, their order-of-magnitude 
and collective behavior could be compared with the pattern and observed in higher-order polynomial fits.

The unnatural behavior of the higher FF derivatives is a consequence of analyticity and the
singularities of the $\pi N$ Born amplitudes, which govern the behavior of the spectral function
in the near-threshold region. Dispersive fits to the low-$Q^2$ FF data correctly implement
these features and can provide reliable results for the higher 
FF derivatives \cite{Belushkin:2006qa,Lorenz:2012tm}.
\subsection{Spacelike EM form factors}
\begin{figure*}
\begin{center}
\epsfig{file=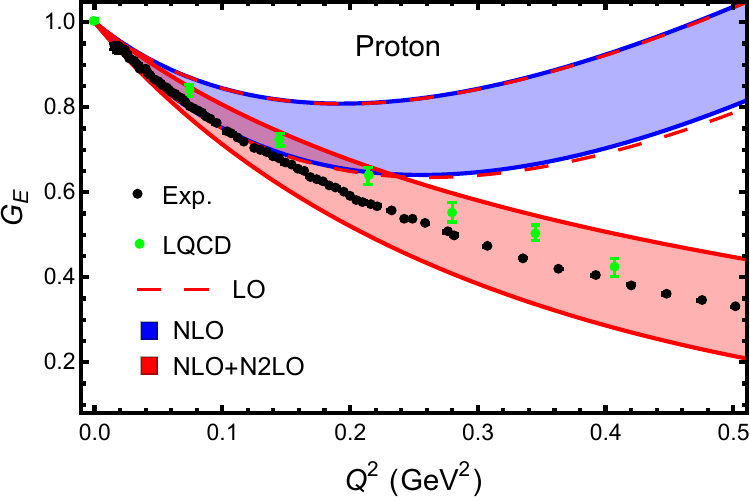,width=.45\textwidth,angle=0} \epsfig{file=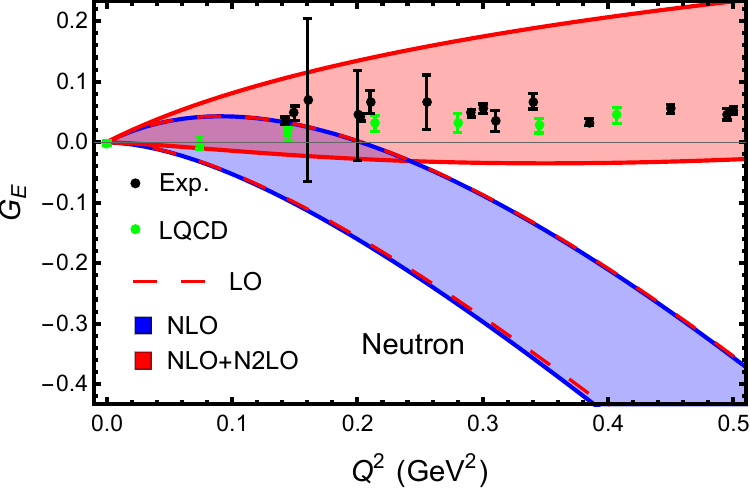,width=.45\textwidth,angle=0} 
\epsfig{file=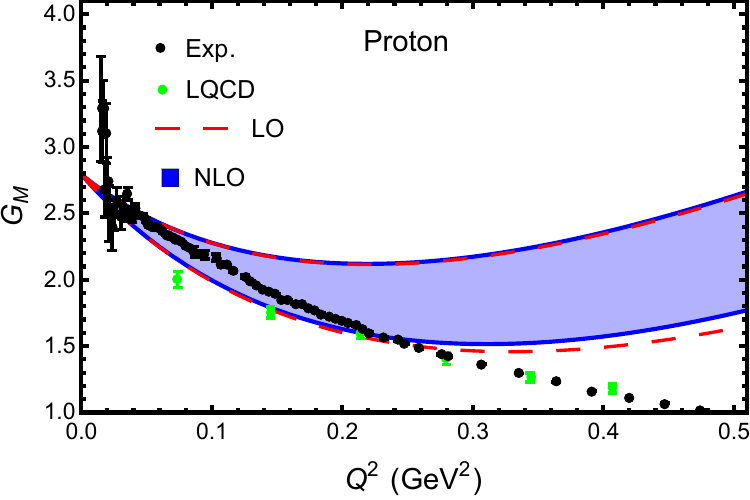,width=.45\textwidth,angle=0} \epsfig{file=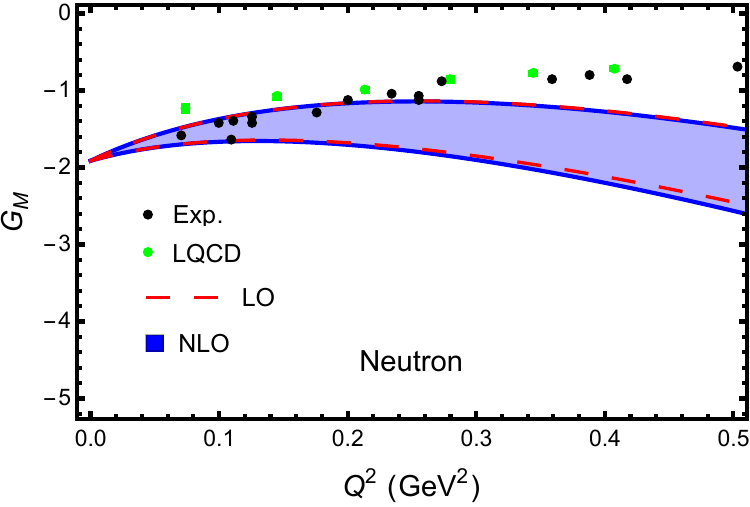,width=.45\textwidth,angle=0} 
\caption{DI$\chi$EFT predictions for the proton and neutron electric and magnetic FFs,
obtained with the twice-subtracted dispersion relation for the isovector FF, Eq.~(\ref{disp_ff_twice}), 
and the empirical parametrization of the isoscalar FF Eq.~(\ref{isoscalar_empirical}).
The results are compared to the experimental data of the 
A1 Collaboration \cite{Bernauer:2010wm,Bernauer:2013tpr} and the LQCD
results from the ETM Collaboration \cite{Alexandrou:2017ypw}.
\label{fig:ff}}
\end{center}
\end{figure*}

The DI$\chi$EFT approach also allows us to calculate the nucleon FFs at finite $t < 0$,
where they are measured in $eN$ elastic scattering experiments. For the isovector FFs we 
use the twice-subtracted dispersion relations
\ba
G_i^V (t) &=& \; G_i^V (0) \; + \; t \, \frac{dG_i^V}{dt} (0) \nonumber \\
&+& \frac{t^2}{\pi} \int_{4M_\pi^2}^\infty dt' \; \frac{\textrm{Im}\, G_i^V (t')}{{t'}^2(t' - t)} 
\hspace{1em} (i=E, M) .
\label{disp_ff_twice}
\ea
Here the FFs at $t = 0$ (charge and magnetic moment) and the first derivatives (electric 
and magnetic radii) are taken as input, and the dispersion relation predicts the 
$t$--dependence starting from the second order. The integrals in Eq.~(\ref{disp_ff_twice})
are well convergent and can be evaluated with the DI$\chi$EFT spectral functions.
For the isoscalar FF we use the empirical parametrization Eq.~(\ref{isoscalar_empirical}),
which imposes the correct value of the FF at $t = 0$. Combining the two we predict the 
individual proton and neutron FFs. Figure~\ref{fig:ff} summarizes the results. 
The estimated uncertainties are dominated by those of the isoscalar component, for which
we have only the empirical parametrization. One observes: (a) In the electric FFs $G_E^p$ and
$G_E^n$, the LO and NLO approximations describe the experimental data only up to $Q^2 \sim 0.1$ GeV$^2$,
while the NLO + N2LO approximations show good agreement with the data up to $Q^2 \sim 0.5$ GeV$^2$.
This reflects the improvement of the isovector electric spectral function due to the partial N2LO
corrections; see Figs.~\ref{fig:Js} and \ref{Fig:ImGEGM}. (b) In the magnetic FFs $G_M^p$ and
$G_M^n$, our results describe the data up to $Q^2 \sim 0.2$ GeV$^2$. In this channel the N2LO 
corrections cannot be estimated using the method of Sec.~\ref{subsec:higher_order}.
Altogether we obtain a very satisfactory description of the nucleon EM FFs with our 
dynamical approach.
%
% FIGURE
%
\section{Summary}
\label{sec:summary}
This work reports a study of the nucleon EM FFs at momentum transfers $|t| \ll 1 \, \textrm{GeV}^2$
using a new method combining $\chi$EFT and dispersion analysis (DI$\chi$EFT). The isovector spectral 
functions on the two-pion cut are constructed through the elastic unitarity condition. The $N/D$ method
is used to separate effects of $\pi\pi$ rescattering from the coupling of the $\pi\pi$ system 
to the nucleon. $\chi$EFT is employed to calculate the real functions $J_\pm^1(t)$ describing
the $\pi\pi$ coupling to the nucleon, which are free of $\pi\pi$ rescattering effects, resulting
in good convergence. $\pi\pi$ rescattering effects are included through the timelike pion FF $|F_\pi(t)|^2$,
which can be taken from $e^+e^-$ annihilation data or LQCD calculations. The new organization 
is consistent with basic principles of $\chi$EFT and represents a major improvement over traditional 
direct calculations of the spectral functions. It allows us to calculate the isovector spectral 
functions up to $t \sim 1\, \textrm{GeV}^2$ (including the $\rho$ meson region) with controlled accuracy.
With these spectral functions we are able to evaluate elements of the FFs (radii, higher derivatives, 
$t$-dependence in spacelike region) using well-convergent subtracted dispersion relations.

The new method permits a realistic description of the low-$t$ nucleon FFs and their derivatives. 
While the basic features of the FFs are rooted in analyticity and have been studied earlier 
in empirical dispersion theory, the new element is that the spectral functions can now be computed 
in a $\chi$EFT framework with controlled accuracy. It makes it possible to represent the information 
content of the nucleon FFs in the form of a few physical masses and LECs, resulting in a significant 
reduction of complexity. It also enables new interpretations of FFs in terms of spatial 
densities \cite{Granados:2013moa,Alarcon:2017asr} and a space-time picture 
of chiral processes \cite{Granados:2015rra,Granados:2015lxa,Granados:2016jjl}.

Our study shows that the derivatives of the EM FFs involve two dynamical scales. 
The first derivative is governed by the scale $1/M_V^2$, while higher derivatives 
are governed by the scale $1/(4 M_\pi^2) \gg 1/M_V^2$ and therefore appear unnaturally large. 
The rich structure attests to the fact that, through analyticity and the dispersion relations,  
the FFs of the nucleon are connected to its hadronic couplings and excitation 
spectrum and reflect the multiple dynamical scales characterizing the latter.
The DI$\chi$EFT calculations provide an explicit realization of this general feature.
While the predicted pattern of the higher moments is likely far beyond what can be extracted
from experimental data, our analytic functions can be used in Monte-Carlo simulations
to study how well the first few moments can be extracted under realistic conditions.
Our numerical estimates of the higher derivatives can be used in an empirical analysis
based on bounded least-squares regression. An analysis incorporating theoretical 
constraints from DI$\chi$EFT is in progress \cite{AHW}.

The DI$\chi$EFT FF calculations described here could be extended in several directions.
The method could be applied to the $N$--$\Delta$ transition FFs as well as the EM 
FFs of the $\Delta$ itself, which are defined rigorously in the context of $S$--matrix 
theory (as poles in the $N \rightarrow \pi N$ and $\pi N \rightarrow \pi N$ EM transition 
amplitudes) and have been studied in relativistic $\chi$EFT \cite{Ledwig:2011cx,Ledwig:2010ya}.
The method could also be applied to nucleon FFs of other $G$-parity-even operators, such
as the energy-momentum tensor or higher moments of the GPDs. Finally, one might contemplate
extending the DI$\chi$EFT approach to nucleon FFs of $G$-parity odd operators with a 3-pion cut, 
using methods of 3-body elastic unitarity that are presently being developed 
for the analysis of meson decays and LQCD calculations \cite{Mai:2017vot,Briceno:2016xwb}. 

The present study demonstrates the potential of $\chi$EFT to yield fully predictive results
for conventional nucleon structure observables. The same approach can be applied to hadronic 
structure elements appearing in searches for Physics Beyond the Standard Model; 
see e.g.~Refs.~\cite{Alarcon:2011zs,Alarcon:2012nr,Alarcon:2013cba}.
\section*{Acknowledgments}
We are indebted to C.~Alexandrou, J.~Bernauer, and A.~Pineda for providing us with 
experimental and theoretical data shown in the tables figures, and to D.~Higinbotham
for valuable discussions and help with the compilation of data and references.

This material is based upon work supported by the U.S.~Department of Energy, 
Office of Science, Office of Nuclear Physics under contract DE-AC05-06OR23177.
This work was also supported by the Spanish Ministerio de Econom\'ia y Competitividad and
European FEDER funds under Contract No. FPA2016-77313-P.
\appendix
\section{$J$ functions in $\chi$EFT}
\label{app:expressions}
In this appendix we list the $\chi$EFT expressions for the real functions 
$J_\pm^1(t)$, Eq.~(\ref{J_pm_def}), which appear in the $N/D$ representation
of the elastic unitarity condition and are used in the analytical and numerical 
studies described in the text. In the following $4 M_\pi^2 < t < 4 m_N^2$, and
\be
k_{\rm cm} = \sqrt{t/4 - M_\pi^2}, \hspace{2em} \widetilde{p}_{\rm cm} = {\textstyle \sqrt{m_N^2 - t/4}},
\label{k_appendix}
\ee
are, respectively, the physical pion CM momentum and the unphysical nucleon CM momentum in the 
$\pi\pi \rightarrow N\bar N$ process. 
The functions resulting from the Weinberg-Tomozawa contact term (Fig.~\ref{fig:eft}b)
and the $N$ Born term (Fig.~\ref{fig:eft}a) are
\ba
J_1^+(t) [\textrm{LO, cont}] &=& \frac{m_N}{24\, \pi f_\pi^2} ,
\\
J_1^-(t) [\textrm{LO, cont}] &=& \frac{\sqrt{2}}{24 \pi f_\pi^2} , 
\ea
\ba
J_1^+(t) [\textrm{LO}, N] &=&
\frac{g_A^2 m_N^3 A_N^2}{16 \pi f_\pi^2 \, \widetilde p_{\rm cm}^{\, 3} \, k_{\rm cm}^3}
(- \arctan x_N + x_N)
\nonumber \\[1ex]
&-& \frac{g_A^2 m_N}{24\, \pi f_\pi^2} ,
\label{J_plus_LO_N}
\\[1ex]
J_1^-(t) [\textrm{LO}, N] 
&=&  \frac{ \sqrt{2} \, g_A^2 m_N^2 A_N^2}{32 \pi f_\pi^2 \, \widetilde p_{\rm cm}^{\, 3} \, k_{\rm cm}^3} 
\nonumber \\[1ex]
&& \times [(x_N^2 + 1) \arctan x_N - x_N]
\nonumber \\[1ex]
&-& \frac{\sqrt{2} \, g_A^2}{24 \pi \, f_\pi^2} ,
\label{J_minus_LO_N}
\ea
\ba
A_N &\equiv& t/2 - M_\pi^2 ,
\label{A_N_def}
\\[1ex]
x_N &\equiv& \frac{2 k_{\rm cm} \widetilde{p}_{\rm cm}}{A_N} 
\nonumber \\[1ex]
&=& \frac{2 \, \sqrt{t/4 - M_\pi^2} {\textstyle \sqrt{m_N^2 - t/4}}}{t/2 - M_\pi^2} .
\label{x_N}
\ea
The contributions of the $\Delta$ Born term (Fig.~\ref{fig:eft}c) are
\ba
J_+^1(t)[\textrm{LO}, \Delta]
&=& \frac{h_A^2 \, A_\Delta (2 \widetilde{p}_{\rm cm}^{\, 2} F - A_\Delta m_N G)}
{192 \pi  f_{\pi }^2 \, \widetilde{p}_{\rm cm}^{\, 3} k_{\rm cm}^3} 
\nonumber \\[1ex]
&& \times (\arctan x_\Delta - x_\Delta) 
\nonumber \\[1ex]
&+& \frac{h_A^2 \, D_{\Delta +}}{432  \pi  f_{\pi }^2 m_\Delta^2} ,
\label{J_plus_LO_Delta}
\\[1ex]
J_-^1(t)[\textrm{LO}, \Delta]
&=& \frac{\sqrt{2} \, h_A^2 \, A_\Delta^2 G}
{384 \pi  f_{\pi }^2 \, \widetilde{p}_{\rm cm}^{\, 3} k_{\rm cm}^3}
\nonumber \\[1ex]
&& \times \left[ (x_\Delta^2 + 1) \arctan x_\Delta - x_\Delta \right]
\nonumber \\[1ex]
&+& \frac{\sqrt{2} \, h_A^2 \, D_{\Delta -}}{864 \pi  f_{\pi }^2 m_\Delta^2} ,
\hspace{1em}
\label{J_minus_LO_Delta}
\ea
\ba
A_\Delta &\equiv& t/2 - M_\pi^2 + m_\Delta^2 - m_N^2 ,
\\[1ex]
x_\Delta &\equiv& \frac{2 k_{\rm cm} \widetilde{p}_{\rm cm}}{A_\Delta} 
\nonumber
\\[1ex]
&=& 
\frac{2 \, \sqrt{t/4 - M_\pi^2} {\textstyle \sqrt{m_N^2 - t/4}}}{t/2 - M_\pi^2 + m_\Delta^2 - m_N^2} .
\label{x_Delta}
\ea
The functions $F$ and $G$ appearing in the first terms of Eqs.~(\ref{J_plus_LO_Delta})
and (\ref{J_minus_LO_Delta}) are \cite{Alarcon:2017ivh}
\ba
F &\equiv& \alpha (m_\Delta + m_N) + \frac{\beta}{3} (m_\Delta - m_N) ,
\label{F_def}
\\[1ex] 
G &\equiv& - \alpha + \frac{\beta}{3} ,
\label{G_def}
\\[1ex] 
\alpha &\equiv& \frac{t}{2} - m_N^2 + 
\frac{(m_\Delta^2 + m_N^2 - M_\pi^2)^2}{4 m_\Delta^2} ,
\\[1ex] 
\beta &\equiv& \left( m_N + \frac{m_\Delta^2 + m_N^2 - M_\pi^2}{2 m_\Delta}\right)^2 ;
\ea
they are the invariant amplitudes of $\pi N$ scattering at $t > 4 M_\pi^2$ and 
$s = m_\Delta^2$ in the conventions of Ref.~\cite{Granados:2016jjl}.
The functions $D_{\Delta +}$ and $D_{\Delta -}$ appearing in the
second terms in Eqs.~(\ref{J_plus_LO_Delta}) and (\ref{J_minus_LO_Delta}) are
\ba
D_{\Delta +}  &=&  m_N^3 + 2m_N^2 m_\Delta + m_N m_\Delta^2 -m_N M_\pi^2 \nonumber\\
                      &+& (m_N - m_\Delta)t , \\
D_{\Delta -}  &=& -10 m_N^2-4m_N m_\Delta + 2 m_\Delta^2 \nonumber \\
                     &-& 2 M_\pi^2 + 5 t .
\ea

The $J$ functions resulting from the $N$ and $\Delta$ Born terms
have logarithmic left-hand cuts starting at
\be
\left.
\begin{array}{ll}
N{:} \hspace{1em}     & t < 4 M_\pi^2 - M_\pi^4/m_N^2 \\[.5ex]
\Delta{:} & t < 4 M_\pi^2 - (m_\Delta^2 - m_N^2 + M_\pi^2)^2 / m_\Delta^2
\end{array}
\right\} .
\vspace{.5ex}
\ee
The singularity results from the intermediate baryon lines going on mass shell
and corresponds to the left-hand cut of the $\pi\pi \rightarrow N \bar N$ PWA.
The singularity is contained in the inverse tangent functions in 
Eqs.~(\ref{J_plus_LO_N}), (\ref{J_minus_LO_N}), (\ref{J_plus_LO_Delta}), and (\ref{J_minus_LO_N}),
which have logarithmic branch points at $x_{N, \Delta}^2 = \pm i$.
The $J$ functions do not have a right-hand cut at $t > 4 M_\pi^2$, in accordance with 
their definition within the $N/D$ method, Eqs.~(\ref{J_pm_def}) and (\ref{N_over_D}).
While the expressions in Eqs.~(\ref{J_plus_LO_N}), (\ref{J_minus_LO_N}), 
(\ref{J_plus_LO_Delta}), and (\ref{J_minus_LO_Delta}) contain prefactors with inverse 
powers of $k_{\rm cm}$, they are in fact regular in the limit $k_{\rm cm} \rightarrow 0$,
because the expressions in parentheses/brackets depending on $x_N$ or $x_\Delta$
vanish in the limit, $x_{N, \Delta} = \mathcal{O}(k_{\rm cm})$. Further properties of the $J$ 
functions are discussed in Ref.~\cite{Alarcon:2017ivh}.

The masses and coupling constants used in evaluating the LO expressions are the standard 
values for the $SU(2)$ flavor group \cite{Alarcon:2017ivh}:
$M_\pi = 139\, \textrm{MeV}, f_\pi = 93 \, \textrm{MeV},
m_N = 939\, \textrm{MeV}, g_A = 1.27$, and $m_\Delta = 1232\, \textrm{MeV}, h_A = 2.85$.

The contributions of the NLO contact term in the $\pi N$ amplitude are 
\ba
J_1^+(t) [\textrm{NLO, cont}] &=& \frac{c_4 t}{24 \pi \, f_{\pi }^2} ,
\\[1ex]
J_1^-(t) [\textrm{NLO, cont}] &=& \frac{\sqrt{2} \, m_N c_4}{6 \pi \, f_{\pi }^2} .
\ea
The value of the LEC $c_4$, determined by the procedure described 
in Sec.~\ref{subsec:higher_order}, is given in Eq.~(\ref{c4_adjusted}).
\section{Higher derivatives of proton electric form factor}
\label{app:derivatives}
For reference we present in Table~\ref{tab:proton_higher_moments} our numerical estimates 
of the higher moments of the proton electric FF, obtained by combining the DI$\chi$EFT 
calculation of the isovector moments with the empirical estimate of the isoscalar 
moments based on Eq.~(\ref{isoscalar_empirical}). While individual higher moments have
little physical significance and cannot realistically be extracted from the data, the 
order-of-magnitude and collective behavior of our results could be compared with the patterns 
observed in fits to FF data \cite{Bernauer:2010wm,Bernauer:2013tpr,Lorenz:2012tm,Lee:2015jqa,%
Higinbotham:2015rja,Griffioen:2015hta,Horbatsch:2016ilr,Sick:2017aor,Bernauer:2010zga}.
%
%
% TABLE
%
\begin{table}
\begin{ruledtabular}
\begin{tabular}{lcccc}
$G_E^p$ &  &  & \\
\hline
 &LO & NLO &  NLO+N2LO\\
  $ \langle r^4    \rangle $           (fm$^4$)    &  (2.09, 2.48) & (2.00, 2.53) & (1.16, 1.70) \\[-.3ex]
  $ \langle r^6    \rangle $           (fm$^6$)    &  (10.8, 11.7) & (10.5, 11.8) & (7.59, 9.00) \\[-.3ex]
  $ \langle r^8    \rangle $ ($10^{2} $ fm$^8$)    &  (1.44, 1.48) & (1.42, 1.49) & (1.21, 1.29) \\[-.3ex]
  $ \langle r^{10} \rangle $ ($10^{3} $ fm$^{10}$) &  (4.21, 4.24) & (4.18, 4.26) & (3.86, 3.94) \\[-.3ex]
  $ \langle r^{12} \rangle $ ($10^{5} $ fm$^{12}$) &  (2.13, 2.13) & (2.12, 2.14) & (2.02, 2.04) \\[-.3ex]
  $ \langle r^{14} \rangle $ ($10^{7} $ fm$^{14}$) &  (1.60, 1.61) & (1.60, 1.61) & (1.55, 1.56) \\[-.3ex]
  $ \langle r^{16} \rangle $ ($10^{9} $ fm$^{16}$) &  (1.66, 1.66) & (1.66, 1.67) & (1.61, 1.62) \\[-.3ex]
  $ \langle r^{18} \rangle $ ($10^{11}$ fm$^{18}$) &  (2.25, 2.25) & (2.25, 2.26) & (2.20, 2.21) \\[-.3ex]
  $ \langle r^{20} \rangle $ ($10^{13}$ fm$^{20}$) &  (3.86, 3.86) & (3.85, 3.87) & (3.79, 3.80) \\[-.3ex]
  $ \langle r^{22} \rangle $ ($10^{15}$ fm$^{22}$) &  (8.15, 8.15) & (8.13, 8.16) & (8.00, 8.03) \\[-.3ex]
  $ \langle r^{24} \rangle $ ($10^{18}$ fm$^{24}$) &  (2.08, 2.08) & (2.07, 2.08) & (2.04, 2.05) \\[-.3ex]
  $ \langle r^{26} \rangle $ ($10^{20}$ fm$^{26}$) &  (6.28, 6.28) & (6.27, 6.28) & (6.18, 6.20) \\[-.3ex]
  $ \langle r^{28} \rangle $ ($10^{23}$ fm$^{28}$) &  (2.22, 2.22) & (2.22, 2.22) & (2.19, 2.20) \\[-.3ex]
  $ \langle r^{30} \rangle $ ($10^{25}$ fm$^{30}$) &  (9.11, 9.11) & (9.09, 9.11) & (8.99, 9.01) \\[-.3ex]
  $ \langle r^{32} \rangle $ ($10^{28}$ fm$^{32}$) &  (4.27, 4.27) & (4.27, 4.28) & (4.22, 4.23) \\[-.3ex]
  $ \langle r^{34} \rangle $ ($10^{31}$ fm$^{34}$) &  (2.28, 2.28) & (2.27, 2.28) & (2.25, 2.26) \\[-.3ex]
  $ \langle r^{36} \rangle $ ($10^{34}$ fm$^{36}$) &  (1.37, 1.37) & (1.37, 1.37) & (1.35, 1.36) \\[-.3ex]
  $ \langle r^{38} \rangle $ ($10^{36}$ fm$^{38}$) &  (9.20, 9.20) & (9.19, 9.21) & (9.10, 9.12) \\[-.3ex]
  $ \langle r^{40} \rangle $ ($10^{39}$ fm$^{40}$) &  (6.88, 6.88) & (6.87, 6.89) & (6.81, 6.83) \\[-.3ex]
\end{tabular}
\caption{Higher-order moments of $G_E^p$, obtained by combining the DI$\chi$EFT 
calculation of the isovector derivatives with an empirical estimate of the isoscalar derivatives.}
\label{tab:proton_higher_moments}
\end{ruledtabular}
\end{table}
\end{document}